\documentclass[twocolumn,runningheads]{svjour2}
\smartqed
\usepackage{graphicx}
\usepackage{mathptmx}   
\usepackage[mathcal]{eucal}
\usepackage{amssymb}
\usepackage{amsmath}
\usepackage[T1]{fontenc}

\usepackage[authoryear]{natbib}

\def\bs {\boldsymbol}
\newcommand{\inv} {\frac {1}}
\def\msol {{\mathrm{M}_\odot}}

\newcommand{\rotbar}{{\bar \Omega}}

\newcommand{\deriv} [2] {\frac {d #1 } {d #2} }

\def\i {{\emph{i}}}

\def\dst {{\displaystyle}}

\def\etao {{\eta_0}}

\def\you {{y_{01}}}
\def\zo {{z_{0}}}
\def\yod{{y_{02}}}
\def\yot{{y_{03}}}
\def\yoc{{y_{04}}}

\def\yu  {{y_1}}

\def\yd  {{y_2}}
\def\yt  {{y_3}}
\def\yc  {{y_4}}

\def\geff {{g_{\mathrm{eff}}}}

\def\sigocar {{\sigma_0^2}}

\def\wua {{\omega_{1,a}}}

\def\wub {{\omega_{1,b}}}

\def\w   {{\omega}}

\def\wo  {{\omega_0}}

\def\wut {{\tilde{\omega}_1}}
\def\wdt {{\tilde{\omega}_2}}

\def\mua {{\mu_{a}}}
\def\mub {{\mu_{b}}}

\def\xig   {{\bs {\xi}}}

\def\xiu {{\bs \xi}_{1}}

\newcommand{\eqn} [1] {
\begin{equation}
#1
\end{equation}}

\newcommand{\eqna} [1] {
\begin{eqnarray}
#1
\end{eqnarray}}

 \newcommand{\filou}{{\sc filou}}
 
 \newcommand{\kms} {\mathrm{km}\,\mathrm{s}^{-1}}
 \newcommand{\Ux}{U_{\chi}}
 
 \newcommand{\soufi}{SGD98}
 \newcommand{\suarez}{SGM06}
 \newcommand{\thesis}{S02}
 \newcommand{\gp}{$gp\,$}

 \newcommand{\gds}{$\gamma$ Doradus stars}
 \newcommand{\ds}{$\delta$ Scuti}
 \newcommand{\dss}{$\delta$ Scuti stars}
 \newcommand{\muHz}{\mu\mbox{Hz}}
\journalname{Astrophysics and Space Science (CoRoT/ESTA Volume)}
\begin{document}

\title{\filou\ oscillation code}

\titlerunning{\filou\ oscillation code}        

\author{J.C. Su\'arez \and
        M.J. Goupil}
\authorrunning{Su\'arez, J.C.}                     

\institute{J.~C. Su\'arez \at 
           Instituto de Astrof\'{\i}sica de Andaluc\'{\i}a (CSIC)\\
           Camino Bajo de Hu\'etor, 50 - CP3004 - Granada, Spain \\
           \email{jcsuarez@iaa.es} \and
	   M.~J. Goupil \at
	   LESIA, Observatoire de Paris-Meudon, UMR8109, Meudon, France\\
	   \email{mariejo.goupil@obspm.fr}	   
}

\date{Received: date / Accepted: date}

\maketitle
\begin{abstract}
The present paper provides a description of the oscillation code 
\filou, its main features, type of applications it can be used for, 
and some representative solutions. The code is actively involved
in CoRoT/ESTA exercises (this volume) for the preparation for the
proper interpretation of space data from the CoRoT mission.
Although CoRoT/ESTA exercises have been limited to the oscillations 
computations for non-rotating models, the main characteristic
of \filou\ is, however, the computation of radial and non-radial 
oscillation frequencies in presence of rotation. In particular, \filou\ 
calculates (in a perturbative approach) adiabatic oscillation frequencies 
corrected for the effects of rotation (up to the second order in the rotation 
rate) including near degeneracy effects. Furthermore, 
\filou\ works with either a uniform rotation or a radial differential
rotation profile (shellular rotation), feature which makes the code
singular in the field.
\end{abstract}
\section{Introduction}\label{intro}
Numerous oscillation codes currently provide oscillation
modes for polytropes and 1D models representative of different kind of pulsating stars.
The history of the oscillation code \filou\ is particularly associated
with \dss. Originally developed by F. Tran Minh and L. L\'eon
at Observatoire de Paris-Meudon \citep[see][]{filou}, the
code has undergone several modifications and improvements
in order to correct the oscillation frequencies for the effects
of rotation. In particular, the inclusion of these corrections
(up to the second order including near degeneracy effects) to the oscillation code
and its numerical tests was part of my PhD. work 
\citep{SuaThesis} (hereafter \thesis). 
In that work oscillation computations were extended to
the case of models including a radial differential rotation profile 
$\Omega=\Omega(r)$, i.e., a radial-dependent differential rotation 
(the so-called \emph{shellular} rotation). This last characteristic is, 
by now, unique, and makes this code singular in the field. 

Although \filou\ is currently optimised for the study of the 
pulsational behaviour of intermediate-mass classical pulsators, 
namely \dss\ and \gds,  the code is of universal use.
It has been used, for instance, to model individual 
\dss\ like the well-known Altair \citep{Sua05altairII}, or 
29~Cygnus \citep{Casas06}, as well as to 
study \dss\ in open clusters \citep{Sua02aa, Fox06, Sua07gammes}; 
Moreover, it has served to model high-amplitude \dss\ \citep{Poretti05hads} 
and to analyse the effect of rotation on Petersen diagrams 
\citep{Sua06pdrot,Sua07pdrotII}. Furthermore,
it is worth highlighting the work by \citet{Sua06rotcel} (from now
on SGM06) which
takes advantage of the code's main feature, i.e., the computation of adiabatic 
oscillation in presence of shellular rotation, to analyse the
effect of such type of rotation on adiabatic oscillation of moderately-fast
rotating \dss. 
Concerning \gds, \filou\ has participated, additionally to other
modelling works, in one of the most 
recent and promising asteroseismic tool for the modelling of such stars,
the Frequency Ratio Method (FRM), developed by \citet{Moya05frmI} and 
\citet{Sua05frmII}. 

From the point of view of the numerics, \filou\ solves full sets
of ODE (Ordinary Differential Equations) in a BVP (Boundary Value
Problem), using a combined Galerkine -- B-splines method which
enhances the numerical precision with which the oscillation 
frequencies are calculated. Furthermore, as explained in the following 
sections, it is possible to easily modify numerous numerical 
parameters in order to adjust the calculation optimally for 
the required model, which makes of \filou\ a highly versatile
code.


\section{The adiabatic oscillations equations and boundary
         conditions\label{sec:eqsbound}}
\filou\ is mainly based on the oscillations equations 
and their perturbations developed in \citet{DG92} and \citet{Soufi98}.
In \thesis\ we describe the second-order perturbation
formalism used, which includes the effects of near degeneracy
and considers the presence of a radial differential rotation
(shellular rotation) profile, as well as its implementation
in \filou. 

The notation followed is similar to that used by many other 
oscillation codes, but adapted to the theoretical development
considered. Although the code works with different calculations
schemes, namely, no rotation, Cowling approximation,
and rotation (uniform and differential), in the present 
document only the most general case is considered, i.e.,
the presence of shellular rotation. In such a case, 
oscillations are computed from the so-called \emph{pseudo-rotating}
models, which, as explained in \thesis, are constructed by modifying 
the stellar structure equations such as to include the spherical symmetric 
contribution of the centrifugal acceleration, by means of an 
effective gravity $g_{\mathrm{eff}}=g-{\cal A}_{c}(r)$
where $g$ and ${\cal A}_{c}(r)$ are the local gravity component 
and the centrifugal acceleration, respectively. 
The effects of the non-spherical components of the 
deformation of the star are included through a perturbation 
in the oscillation equations. For instance, the perturbation 
of the mean density of a pseudo-rotating model $\rho_0$ is 
considered of the form $\rho_2 = p_{22}(r)\,P_{2}(\cos\theta)$,
where $p_{22}(r)$ is defined in \suarez\ (Eq.~15).

Furthermore, when near degeneracy is taken into account, the eigenfrequency and 
the eigenfunction of a near-degenerate mode are then assumed of the form:
\eqna{\w^{\mathrm{d}}&=&\bar \omega_{0}+\wut+\wdt \label{eq:defw}\\
      \xig &=&\sum_{j=a,b}\alpha_j (\xi_{0,j}+\xi_{1,j})\,.
\label{eq:defxi}}
where $\bar\omega_0= (\omega_{0,a}+\omega_{0,b})/2$ {\bf and
$\alpha_j$ represent the coefficients of the linear combination
between the two considered degenerate modes.}
Subscripts $a$ and $b$ represent whatever two rotationally coupled
modes. First- and second-order corrections to the eigenfrequency in presence of near 
degeneracy are represented by $\wut$ and $\wdt$ respectively; 
$\xi_{0,j}$ and $\xi_{1,j}$ are the non perturbed and first-order
(see definitions and details in \thesis\ and \suarez).
The computation of individual $\omega_{0,j}$ as well as the
corresponding zeroth- and first-order eigenfunctions 
is described in the next sections.
\begin{figure*}
\centering
   \includegraphics[width=8.5cm]{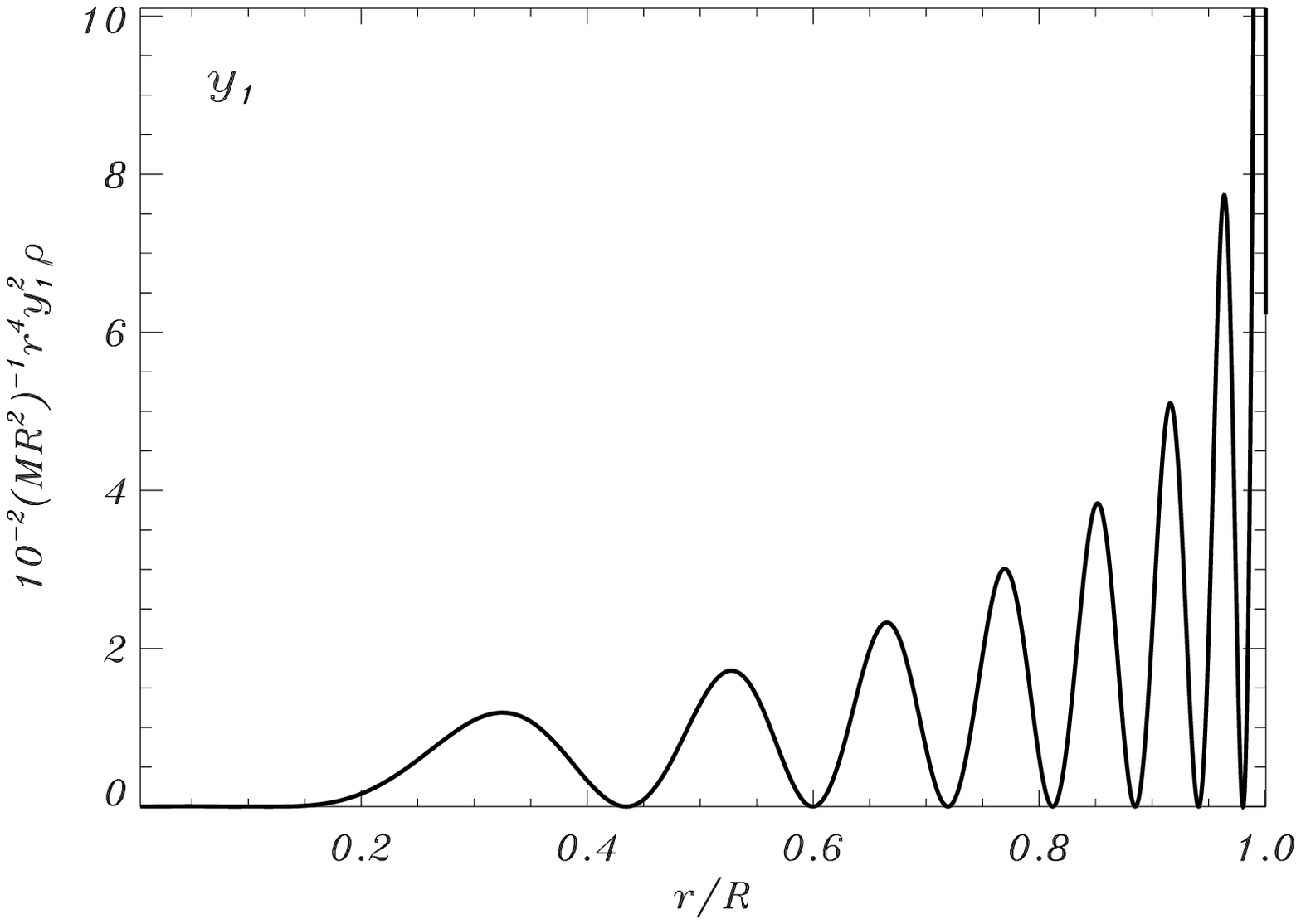}
   \includegraphics[width=8.5cm]{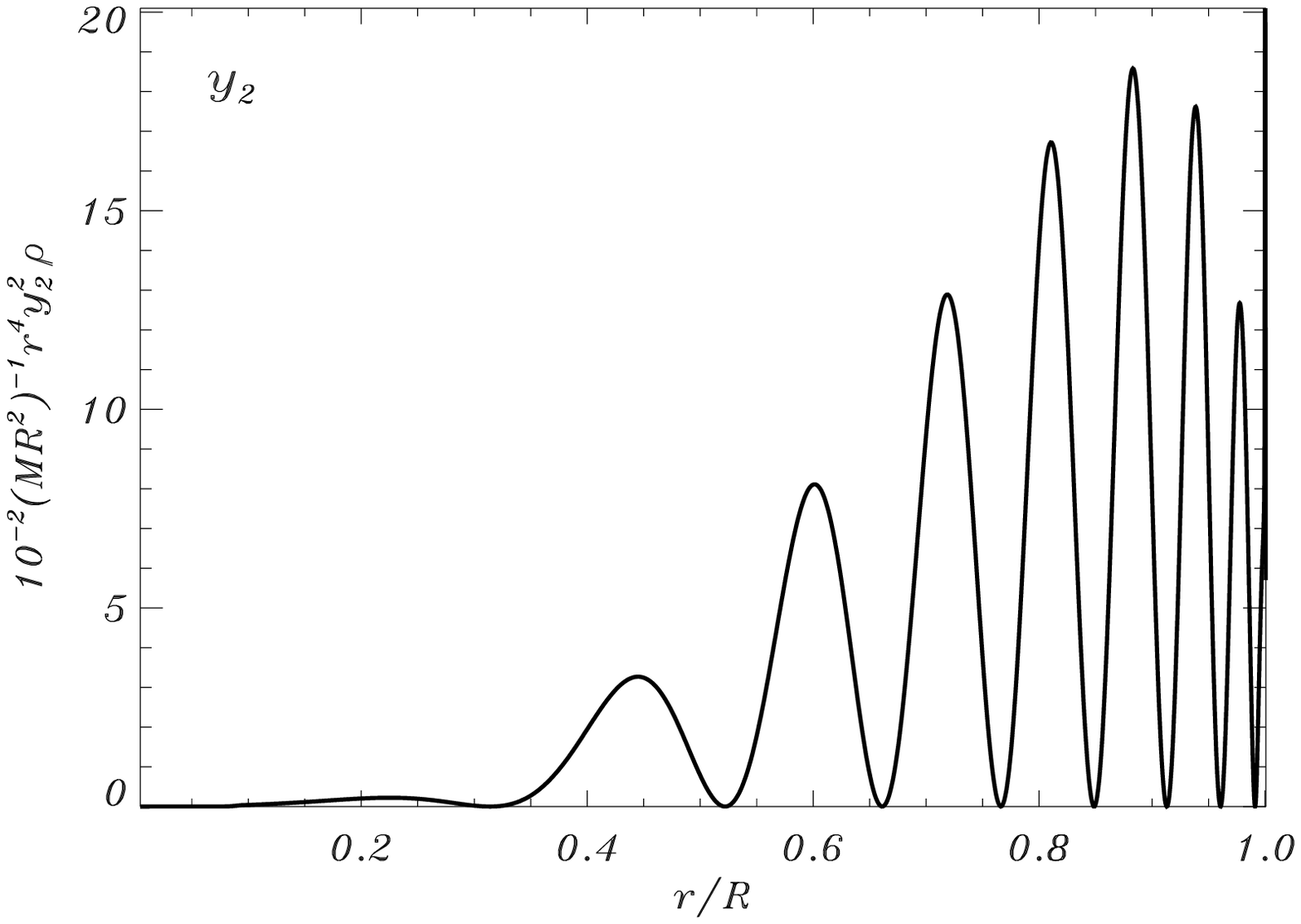}  
   \includegraphics[width=8.5cm]{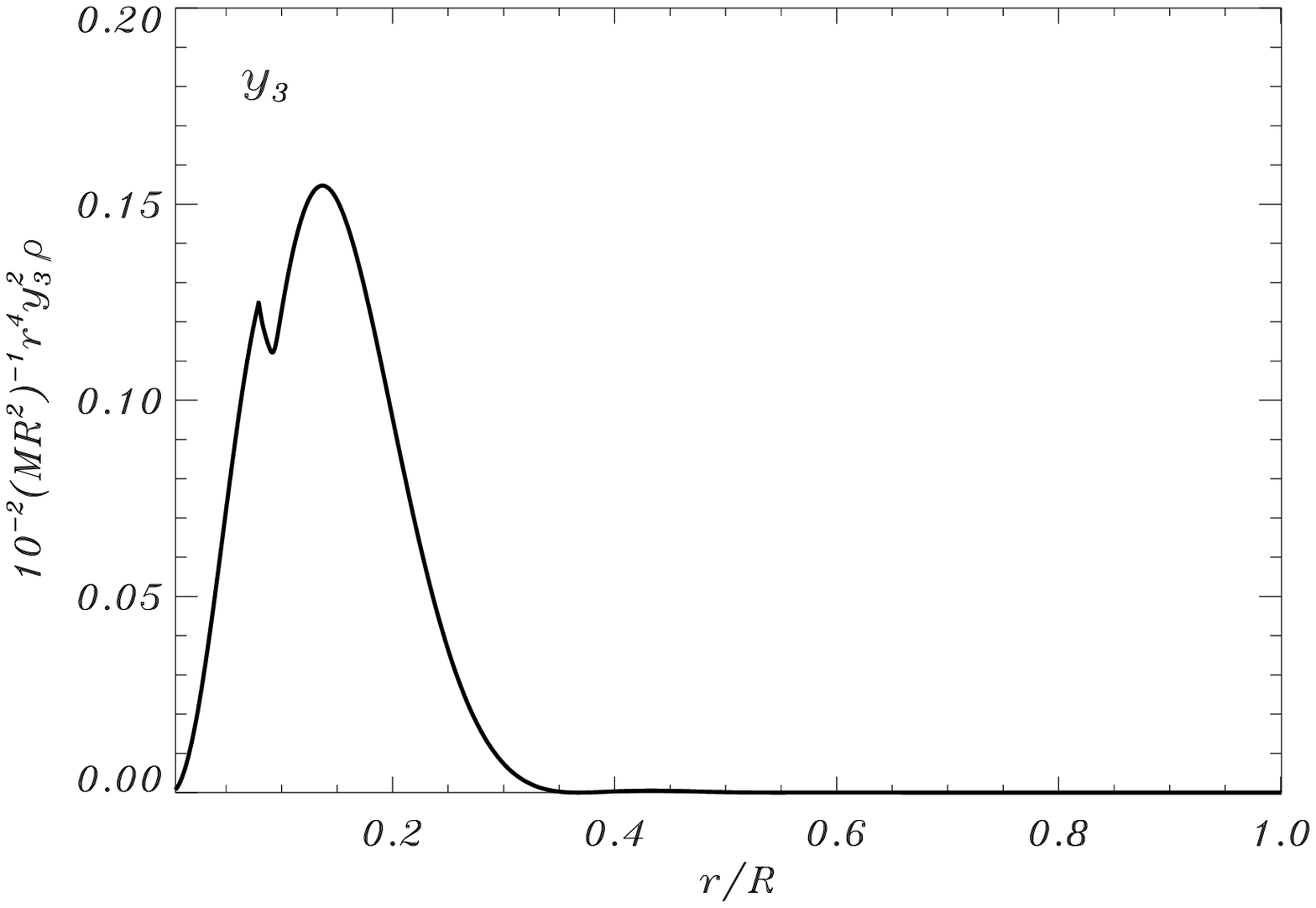}  
   \includegraphics[width=8.5cm]{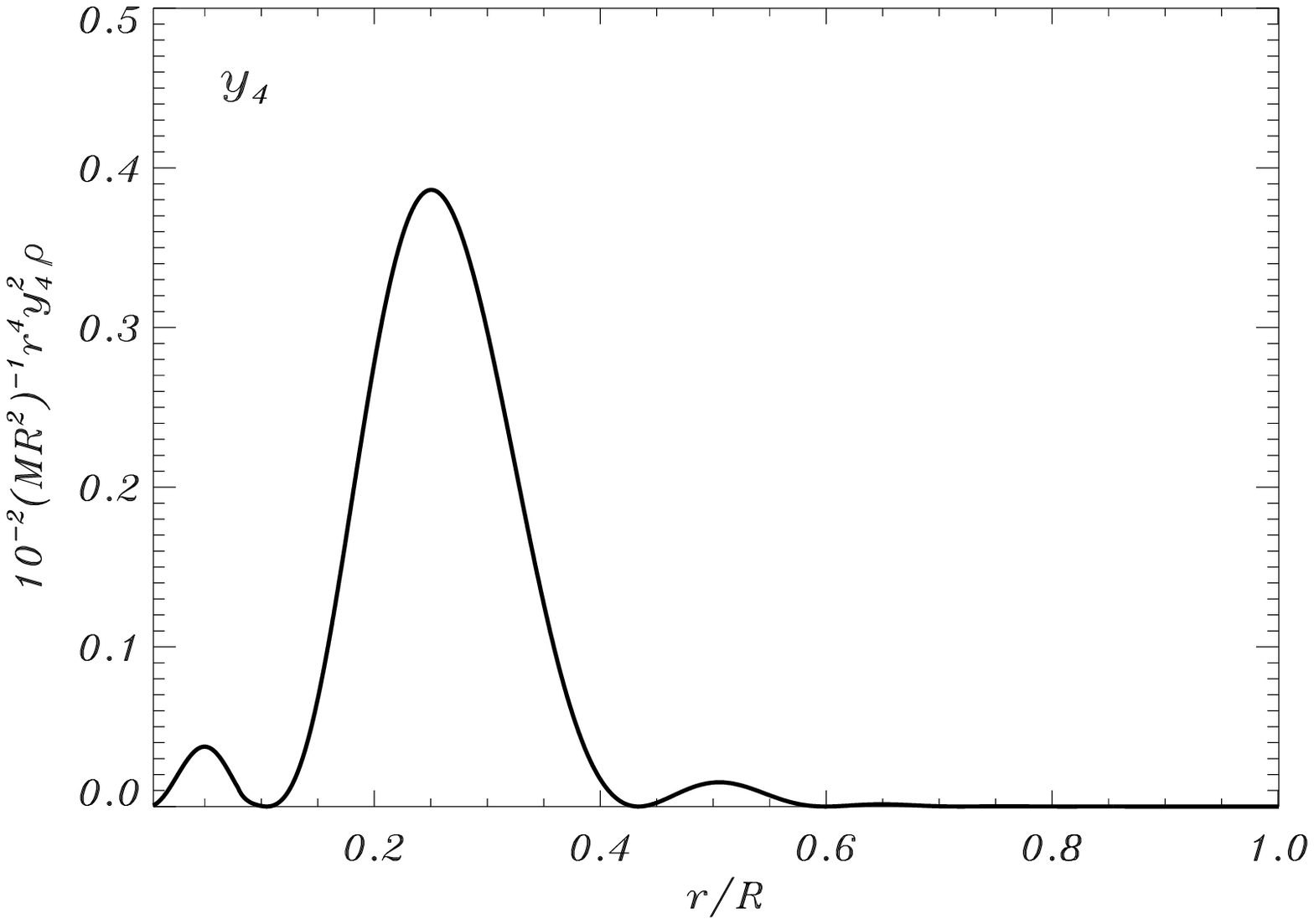}     
   \caption{Normalised eigenfunctions $\you$, $\yod$, $\yot$, and $\yoc$,
            as a function of the normalised radial distance $r/R$, corresponding
	    to the oscillation mode ($n=8,\ell=1$) calculated from a 
	    $1.8\,\msol$ \ds\ star model, with a rotational velocity of
	    $100\,\kms$ at the stellar surface.}
   \label{fig:FFPP}
\end{figure*}
\subsection{Oscillation frequencies of a pseudo-rotating model
           \label{ssec:SYS}}
In order to compute the oscillation frequencies of a pseudo-rotating
model, $\omega_{0,j}$ the following dimensionless quantities are used
\eqna{\you&=&\frac{\xi_r}{r}\label{ap.eq:defy01}, ~~~~~
      \yod=\inv{\geff \,r}\big(\phi\prime
                +\frac{p\prime}{\rho}\big)\label{ap.eq:defy02}\\
      \yot&=&\frac{\phi\prime}{\geff \,r}\label{ap.eq:defy03},~~~~~
      \yoc=\inv{\geff}\deriv{\phi\prime}{r}\label{ap.eq:defy04}}
where $\geff$ represents the effective gravity defined in 
the previous section. {\bf The quantities $p\prime$ and $\phi\prime$,
represent the eulerian perturbation of the pressure and
the gravitational potential (definitions of individual terms can 
be found in \filou\ and \suarez)}.
Considering a differential rotation profile of the form:
\eqn{\Omega(r)= \rotbar \, \dst[1+\eta_0(r)]\,, \label{eq:defeta0}}
where $\rotbar$ represents the rotation frequency at the stellar surface,
the eigenfrequencies (zeroth order) of a pseudo-rotating model are
calculated from the linearised eigenvalue system:
\eqna{x\deriv{\you}{x}&=&\lambda-3\you+\frac{\Lambda}{C_r\sigocar}\yod \nonumber \\                              
      x\deriv{\yod}{x}&=&(C_r\sigma^2_0-A^*)\you+(A^*+1-\Ux)\yod-
                                  A^*\yot \nonumber \\ 
      x\deriv{\yot}{x}&=&(1-U_\chi)\yot+\yoc \label{ap.eq:sist_ordzero}  \\                                    
      x\deriv{\yoc}{x}&=&\frac{U}{1-\sigma_r}
                            \Big[A^*\you+V_g(\yod-\yot)\Big]
                            +\Lambda\yot
			    -\Ux\yoc\,, \nonumber }
which is solved by \filou\ using the dimensionless variable		    				 
$x=r/R$ and $R$ the stellar radius. In the calculation the frequency
is expressed in terms of the dimensionless squared frequency
\eqn{\sigma^2_0 = \frac{\omega^2}{GM/R^3}\label{eq:defsigma}}
as well as other adiabatic quantities (see Appendix in \suarez)
\eqna{A^*\!\!&=& \inv{\Gamma}\deriv{\ln p}{\ln r}-
               \deriv{\ln \rho}{\ln r}\label{ap.eq:defA}\\
      V&=&-\deriv{\ln p}{\ln r},~~~~~~~~~~V_g= \frac{V}{\Gamma_1}\label{ap.eq:defV}\\
      U &=&\deriv{\ln M_r}{\ln r.}\label{ap.eq:defU}}
As in \soufi, the following variables are also employed
\eqna{C &=& \Big(\frac{r}{R}\Big)^3 \frac{M}{M_r},~~~~~~~~~
      C_r = \frac{C}{1-\sigma_r},~~~~~~~~ 
      \sigma_r\!\! = \frac{{\cal A}_c}{g} \\
      \chi &=& \frac{{\cal A}_c}{\geff}  
            \Big(U-3+\deriv{\Omega^2/\bar \Omega^2}{r}\Big)\label{ap.eq:defChi},
	    ~~~~~~~Ux\! =  U+ \chi\,.        \label{ap.eq:defVx}\\
      \lambda &=& V_g(\you-\yod+\yot)\,~~~~~~~\Lambda=\ell(\ell+1),\label{ap.eq:deflambda}}
where $M$ and $m_r$ are the stellar mass and the 
mass enclosed in the sphere of radius $r$, respectively.
\subsection{The boundary conditions\label{ssec:bcond}}

The system above (Eqs.~\ref{ap.eq:sist_ordzero}) is solved with the appropriate
boundary conditions
\eqna{&\!\!\!\!\yod+\you\dst\frac{\dst 3}{\dst V_g}=0,~&
        3\you+\yoc=0~~~(\ell=0)\\
       &\you-\yod\dst\frac{\dst \ell}{\dst C_r\sigma^2_0}=0,~&
        \yoc-\ell\yot=0~~~(\ell\neq0)	
      \label{ap.eq:condlim_center}}
at the centre of the star and,
\eqna{&&\you = 1 \\
      &&\yoc+(\ell+1)\yot= 0  \\
      &&\you\Big(1+\dst\frac{\Lambda}{VC\sigma_0^2}-
      \dst\frac{4+C\sigma^2_0}{V}\Big)-
          \yod\big(1-\dst\frac{\Lambda}{VC\sigma_0^2}\big)+\nonumber\\
      &&  \yot\Big(1+\dst\frac{\ell+1}{V}\Big)=0
       \label{ap.eq:condlim_surf}} 
at the stellar surface. 
Figure~\ref{fig:FFPP} illustrates the solutions for the normalised
eigenfunctions $\you$, $\yod$, $\yot$, and $\yoc$, corresponding to 
a non-radial mixed mode ($n=8,~\ell=1$) obtained from a $1.8\,\msol$
model with a surface rotational velocity of $100\,\kms$, constructed
with 2100 mesh points.

\subsection{First-order perturbed eigenfunctions \label{sap:1stordEigenf}}
When near-degeneracy effects are considered, first-order corrections to
the eigenfunctions are required (see Eq.~\ref{eq:defxi}).
Considering dimensionless variables equivalent to 
Eq.~\ref{ap.eq:defy01}--\ref{ap.eq:defy04} with first-order perturbed 
quantities ($\xi_{1,r}$, $\phi_1^\prime$, $p_1^\prime$), and the zeroth-order 
solutions obtained from Eq.~\ref{ap.eq:sist_ordzero}. \filou\ calculates
such first-order perturbed eigenfunctions solving the following system:
\eqna{x \deriv{\yu}{x} & = & \lambda_1-3\yu + \frac{\Lambda}{C_r\sigocar}\yd \\
			    &+&(\you+\zo)(1+\etao)-(\etao+ \sigma_1) \Lambda \zo 
			    	\nonumber \\
      x \deriv{\yd}{x} & = & (C_r\sigocar-A^*)\yu+(A^*+1-\Ux)\yd-A^*\yt\nonumber\\
                        & + & (\sigma_1+\eta_0)\you -(1+\eta_0) z_0\nonumber\\
      x \deriv{\yt}{x} & = & (1-\Ux)\yt+\yc \label{ap.eq:sist_ord1} \\
      x \deriv{\yc}{x} & = &  \frac{U}{1-\sigma_r} \Big[ A^* \yu + V_g \yd - V_g \yt \Big]
			    +\Lambda \yt -\Ux\yc \nonumber}
where $\lambda_1= V_g(y_1-y_2+y_3)$ and $ \zo = \yod/C\sigocar$.
The horizontal component of $\xiu$ can be written as follows:
\eqn{z_{1}=\frac{y_2}{C\sigma_0^2} + \frac{1+\etao}{\Lambda}y_{01}
              +\left(\frac{1+\etao}{\Lambda}-{ \sigma_1}\right)z_{0},
	      \label{ap.eq:defz1}}
where ${\sigma_1}=C_L-J_0$ represents the first-order correction
of the corresponding eigenfrequency. 

\subsection{The first- and second-order frequency corrections \label{ssec:freqcorr}}
Second-order frequency corrections in
presence of near degeneracy (Eq.~\ref{eq:defw}) are coded in
\filou\ using the following equations (\thesis, \suarez)
\eqn{\wdt= \Big(\frac{\mub+\mua}{2}\Big)\pm\dst\sqrt{{\cal H}_{2,ab}^{(2)}}\,,
\label{eq:solsyscase22}}
in the case that $\delta\omega_0=\omega_{0,a}-\omega_{0,b}$ is $O(\Omega^2)$,
and 
\eqn{\wdt= \Big(\frac{\nu_b+\nu_a}{2}+\frac{\mub+\mua}{2}+ 
\frac{\delta \omega_0^2}{8\bar \omega_0}\Big)\pm
\dst\sqrt{{\cal H}_{2,ab}^{(1)}}\,,
\label{eq:solsyscase12}}
if $\delta\omega_0$ is $O(\Omega)$. In both cases first-order near degeneracy 
effects are implicit, which are also calculated by \filou\ using
\eqn{\wut=\frac{\wua+\wub}{2}\pm\dst\sqrt{{\cal H}_{1,ab}}\,,\label{eq:solwuto1}}
Definitions of all terms involved are given in \suarez. The $\nu$ and $\mu$ variables
contains the corrections (up to the second order) for the effect of
rotation on individual eigenfrequencies $\omega_{0,j}$ obtained from 
Eq.~\ref{ap.eq:sist_ordzero}. Such corrections are coded in \filou\ 
using the Saio's notation,
\eqn{\omega_{j,2}=\omega_{0,j}+\rotbar(C_{\mathrm{L}}-1-J_0)+
              \frac{\rotbar^2}{\omega_{0,j}}\Big(D_0+m^2\,D_1\Big)\label{eq:wsaio}}
where $D_0\equiv X_1+X_2$ and $D_1\equiv Y_1+Y_2$. Definitions and
details can be found in (\suarez, Eqs~8-15).

\section{Structure \& computation schemes\label{sec:STRU}}

\filou\ is composed by a main program and some modules written in {\sc c},
and two subroutines written in {\sc fortran} (77, 95), which
read input data from the equilibrium models and calculate the near 
degeneracy effects for the rotationally coupled modes.

Computation of radial and non-radial oscillation frequencies of a given resonant
cavity (input equilibrium model) is divided into three sequential steps: first, 
zeroth-order oscillation frequencies (eigenvalue, $\wo$) are
computed as described in Sect.~\ref{ssec:SYS}; then, for each eigenfrequency,
the corresponding second-order frequency corrections (without including
near degeneracy effects) are calculated (see Sect.~\ref{ssec:freqcorr}). 
Finally, the code selects, following certain rules (see \thesis\ or \suarez), 
the rotationally coupled modes (only pairs of coupled modes are considered) and
calculates their corresponding near degeneracy correcting terms 
(also described in Sect.~\ref{ssec:freqcorr}).

\subsection{\filou\ inputs \& outputs \label{ssec:IO}}

\filou\ inpunts are essentially some physical quantities
(see Sect.~\ref{ssec:SYS}) which are read from the equilibrium model and some initial 
parameters. Currently, the most updated version of the code allows the use of the 
following input models:
\begin{itemize}
   \item[$\bullet$] {\sc cesam}-type models: v3.*, v4.*, v5.* and 2k
   \item[$\bullet$] {\sc geneve}-type models 
\end{itemize}
The input parameters,  which are set by the user in a text file ({\sc ascii}), are read by 
the code when executed. The main parameters are:
\begin{itemize}
   \item[$\bullet$] The input equilibrium model file.
   \item[$\bullet$] Type of computation. This option allows the user to \emph{force} some
                    kind of computing regime (for instance, Cowling approximation, no rotation,
		    uniform rotation, differential rotation). As well, the user can choose
		    the type of output files required, for instance (only the list of
		    frequencies, include or not the corrections for the effect of rotation,
		    near degeneracy effects, or even the eigenfunctions).
   \item[$\bullet$] Frequency domain and spherical degree $\ell$ range.  
   \item[$\bullet$] Type of boundary conditions (finite/infinite $V$, see 
                    Sect.\ref{ssec:SYS})  
   \item[$\bullet$] Type of node assignation (zeros of $\you$ or JCD method).		      
\end{itemize}
The basic outputs provided by \filou\ are the list of eigenfrequencies and eigenfunctions.
However, it is possible to obtain
output files containing intermediate calculation data.
\begin{figure}
\centering
   \includegraphics[width=8.5cm]{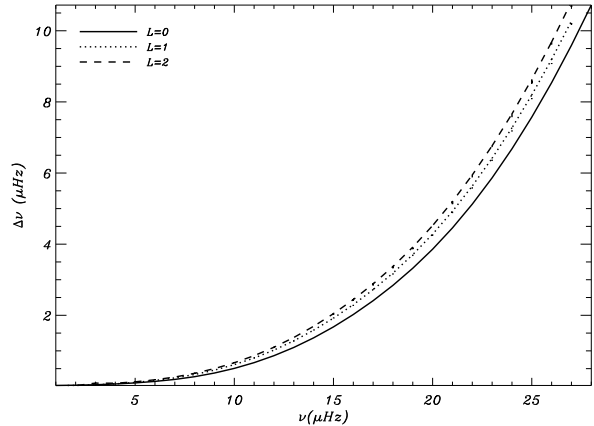}
   \caption{Effect of the number of mesh points on the 
            adiabatic oscillation spectrum of a typical \ds\ star model
	    ($1.8\,\msol$) as a function of the radial order $n$.
	    Curves represent the frequency differences
	    $\Delta\nu=|\nu_{2000}-\nu_{600}|$ obtained 
	     with 600 and 2000 mesh points. }
   \label{fig:ns}
\end{figure}

\section{Numerical techniques\label{sec:NUM}}
The numerical technique followed by \filou\ is based on the Galerkin method,
together with a finite sequence of B-splines, which is characterised
for its flexibility, efficiency and robustness. Although the code
was conceived to solve the numerical problem of stellar non-radial
oscillations, it actually provides solution to any
non-linear system of functional equations, and covers several
specific cases, such as the method of finite elements, Lagrangian 
and/or hermititian of any order, or even the Crank-Nicholson
method of finite differences.

Systems~\ref{ap.eq:sist_ordzero} and \ref{ap.eq:sist_ord1}
are solved by approximating the eigenfunctions (Eqs.~\ref{ap.eq:defy01}-\ref{ap.eq:defy04})
with B-Spline functions. The order of such B-Splines functions can
be chosen by the user, although optimum results are obtained typically
for an optimal order between 4 and 6. The coefficients for each function 
are computed by integration following the Galerkin method. Firstly,
the code scan the user-specified frequency range to obtain a first
\emph{guess} for the eigenfrequencies. Then, \emph{exact} eigenfrequencies
are searched using either the technique of dichotomy or the 
Newton-Raphson method.

\begin{figure}
\centering
   \includegraphics[width=8.5cm]{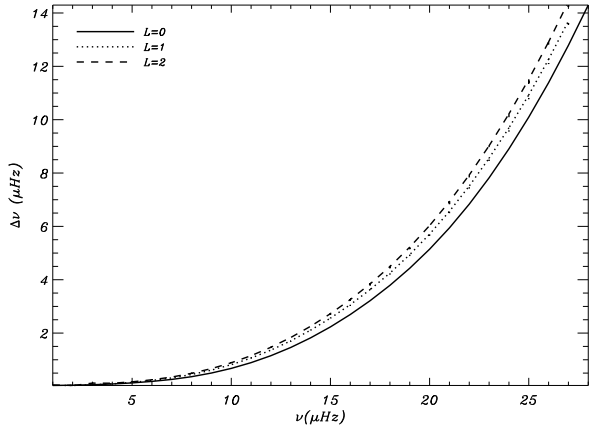}
   \caption{Illustration of the impact of applying Richardson
           extrapolation on the oscillation frequency spectrum
	   computations for the two models used in Fig.~\ref{fig:ns}. 
	   The effect is shown through the frequency differences
	   $\Delta\nu=|\nu_{\mathrm{RE}}-\nu|$, where $\nu_{\mathrm{RE}}$ 
	   represents the oscillation frequencies obtained applying the 
	   Richardson extrapolation.}
   \label{fig:RE}
\end{figure}

It is worth highlighting the numerical versatility of the
code, which can be optimised for the oscillations computation
of very different pulsating stars, g.e., solar-like pulsators (implying
high-order $p$ modes), $g$-mode pulsators (\gds, white dwarfs),
\dss, etc. This is so due to the numerous numerical parameters
that can be adjusted. To name a few, the precision of the 
solutions (zeroth- and first-order eigenfrequencies and eigenfunctions)
required, the size of the internal frequency interval in which
the eigenfrequencies are searched for (this optimises the 
calculations in the cases of high-order and low-order frequencies),
the threshold for valid solutions, etc.
\begin{figure}
\centering
   \includegraphics[width=8.5cm]{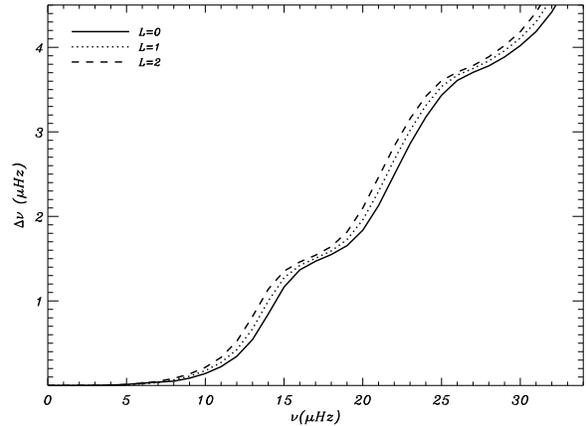}
   \caption{Illustration of the relative difference in frequency obtained,
           for a given main-sequence, \ds\ star model  ($1.8\,\msol$),
	   when using $V\rightarrow 0$ or $V\rightarrow\infty$ in
	   the outer boundary conditions (see Sect.~\ref{ssec:SYS}).}
   \label{fig:BC}
\end{figure}

\subsection{Numerical tests \& results\label{ssec:RES}}
In this section we report succinctly some of the numerical
tests carried out on the code for its optimisation, as well
as some of the tests carried out in the framework of the ESTA 
exercises. 

As has been shown during the ESTA exercises some of the most 
limiting aspects, from the  numerical point of view, in the
calculation of adiabatic oscillations are: the number
of mesh points (including Richardson extrapolation), their
distribution (rezonning), and the type of boundary conditions
used. The current version of \filou\ does not take neither Richardson
extrapolation nor rezonning into account (included in the
next release of the code). Nevertheless we have carried out
some numerical tests of such effects, which are illustrated
in Figs.~\ref{fig:ns}--\ref{fig:BC}, for a typical main-sequence, 
\ds\ star model ($\sim1.8\,\msol$). Notice that all the
effects cause a shift in the oscillation frequency which
increases as far as the radial order increases. This behaviour
is exponential when varying the number of mesh points (Fig.~\ref{fig:ns})
and when using Richardson extrapolation approximation (Fig.~\ref{fig:RE}).
In both cases, for high-order $p$ modes, the effects 
can reach up 10 and $14\,\muHz$, respectively. For the frequency
domain of \dss, i.e. $n\lesssim15$, such effects can
reach up 1 and $2\,\muHz$ which are nonnegligible compared with
the high-precision in frequency detection that CoRoT mission
is expected to provide.

\begin{figure*}
\centering
   \includegraphics[width=8cm]{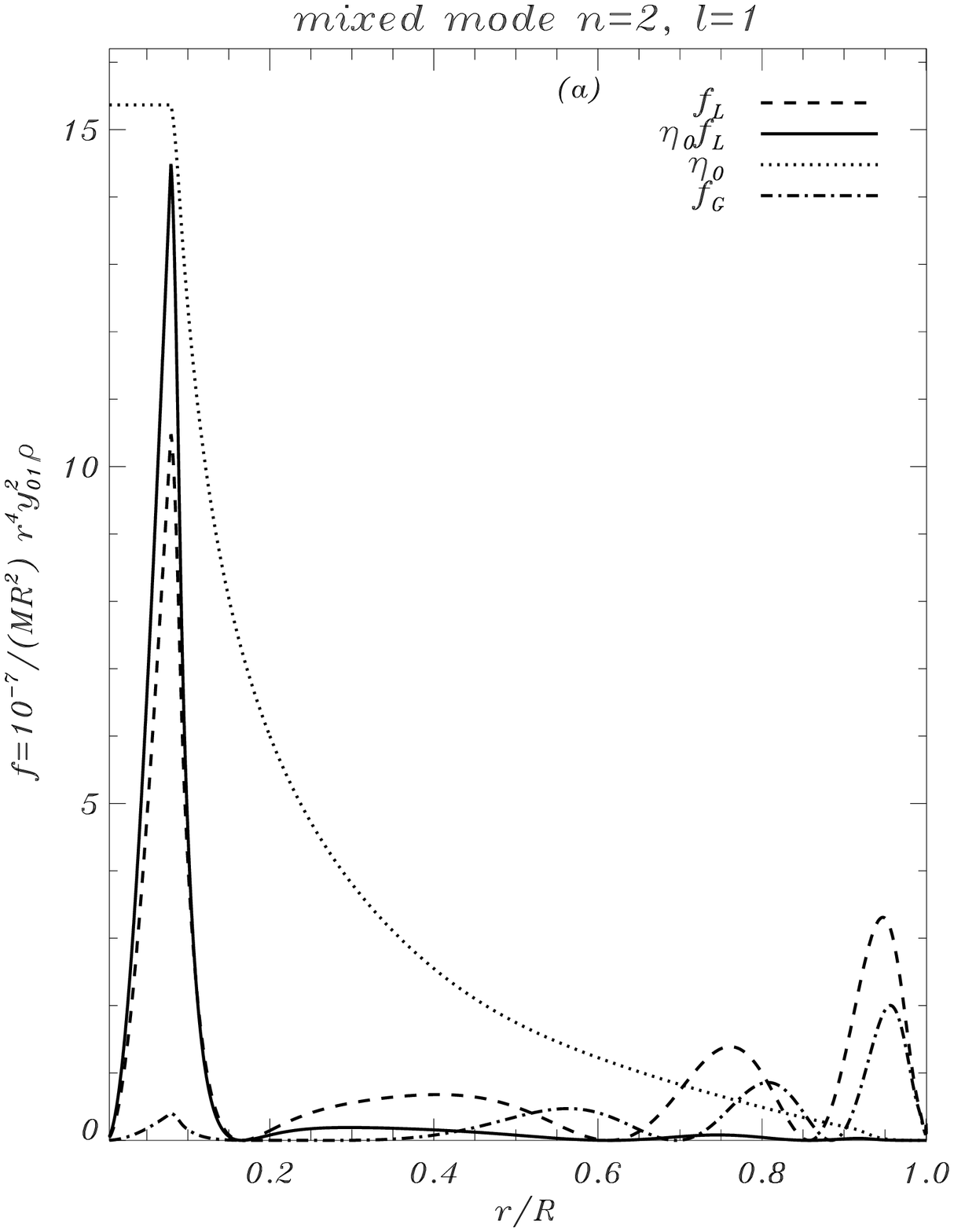}  
   \includegraphics[width=8cm]{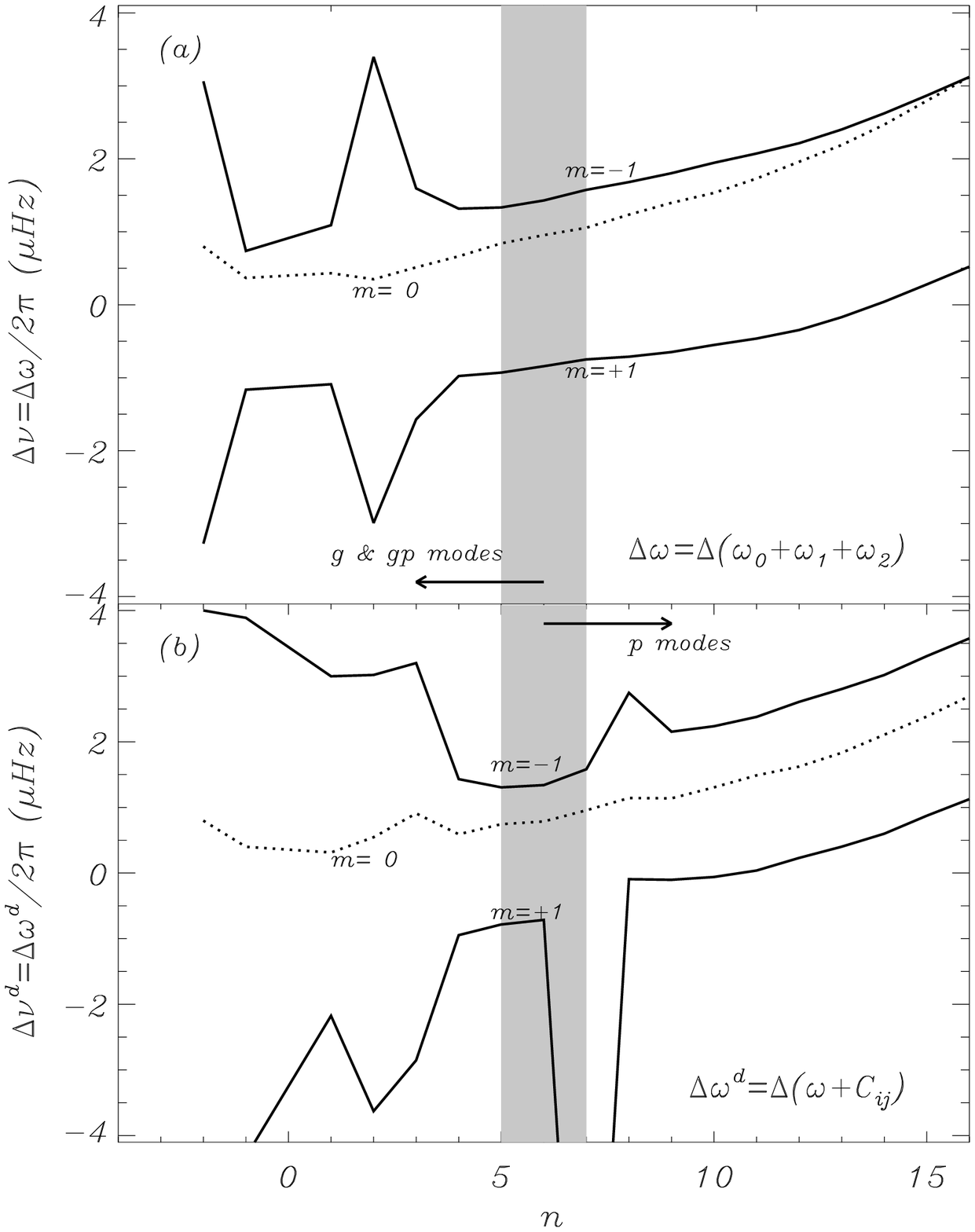}
   \caption{Left panel displays weighted radial displacement eigenfunctions for a mixed mode           
	   as a function of the normalised radial distance $r/R$.
	   Solid and dashed lines represent the $f$ function computed
	   for a shellular-rotating model. Dash-dotted lines represent the $f$ function
           computed for a uniformly-rotating model. Dotted lines represent 
	   the rotation profile (scaled in the figure for clarity) 
	   given by the radial function $\eta_0(r)$.
	   Right panel shows mode-to-mode frequency differences between differentially and
           uniformly-rotating $1.8\,\msol$ models. Symmetric, solid-line branches 
	   represent from top to bottom, differences for $m=-1$ and $m=+1$ 
	   mode frequencies respectively. For $m=0$ modes, differences are 
	   represented by a dotted line. The shaded region 
	   represents an indicative frontier between the region of $g$ and 
	   \gp\ modes (left side) and $p$ modes (right side).
	   Taken from \suarez.}
   \label{fig:SHROT}
\end{figure*}

\subsection{Results. The effect of shellular rotation on adiabatic oscillations}

One of the most important works carried out using the oscillation code \filou\
is the study of the effect of shellular rotation on adiabatic oscillations
(\suarez). Indeed, that work takes advantage of the main feature of 
the code, i.e., the calculation of oscillations in presence of a radial
differential rotation.
Figure~\ref{fig:SHROT} (left panel) illustrates the effect of a shellular rotation on the
radial displacement eigenfunction $\you$, for a \emph{mixed} mode obtained
for a $1.8\,\msol$, \ds\ star model. Such eigenfunctions can be obtained
with \filou\ when calculating the oscillation spectra of 
pseudo-rotating models (see \thesis) computed assuming local conservation 
of the angular momentum. 
Furthermore, the effect of shellular rotation on the oscillation frequencies 
is also significant for \dss. {\bf In Fig.~\ref{fig:SHROT}, such an effect is depicted as 
a function of the radial order. In that figure $\Delta\omega^{\mathrm{d}}$ and 
$\Delta\omega$ represent mode-to-mode frequency differences with (bottom panel)
and without (top panel) taking the effect of near degeneracy into account.
These quantities are calculated, for a given mode, as 
the difference between the oscillation frequency obtained assuming shellular 
rotation with the oscillation frequency obtained assuming uniform rotation.
The term $C_{ij}$ represents the additional effect of near degeneracy on the oscillation 
frequencies for two degenerate modes $i$ and $j$ (see \suarez\ for more details).}
As shown in \suarez, for $g$ and mixed modes, significant effects (up to $3\,\muHz$)
on the the oscillation frequencies are predicted. For high-frequency $p$ modes, such effects 
can reach up to $1\,\muHz$. Such effects are likely to be detectable with CoRoT data, provided numerical 
eigenfrequencies reach the level of precision required. See \suarez\ for a detailed
discussion on these results.

\begin{acknowledgements}   
    JCS acknowledges support at the Instituto de 
    Astrof\'{\i}sica de Andaluc\'{\i}a (CSIC) by an I3P contract
    financed by the European Social Fund and also acknowledges
    support from the Spanish Plan Nacional del Espacio under 
    project ESP2004-03855-C03-01.
\end{acknowledgements}

%
%
%

\end{document}